\documentstyle[epsf,aps]{revtex}

\def\be{\begin{equation}}
\def\ee{\end{equation}}
\def\bea{\begin{eqnarray}}
\def\eea{\end{eqnarray}}

\newcommand{\stac}[2]{\stackrel{\scriptscriptstyle {#1}}{#2}}

\begin{document}

\twocolumn[\hsize\textwidth\columnwidth\hsize\csname
@twocolumnfalse\endcsname

\title{Effective theory for close limit of two branes}
\author{Tetsuya Shiromizu$^{(1,3)}$ Kazuya Koyama$^{(2)}$ and Keitaro Takahashi$^{(2)}$}

\address{$^{(1)}$ Department of Physics, Tokyo Institute of Technology, Tokyo 152-8551, Japan}

\address {$^{(2)}$ Department of Physics, The University of Tokyo, Tokyo 113-0033, Japan}

\address{$^{(3)}$ Advanced Research Institute for Science and Engineering,
Waseda University, Tokyo 169-8555, Japan}

\date{\today}

\maketitle

\begin{abstract}
We discuss the effective theory for the close limit of two branes in 
a covariant way. To do so we  
solve the five dimensional Einstein equation along the direction of the extra dimension. 
Using the Taylor expansion we solve the bulk spacetimes and derive the effective theory 
describing the close limit. 
We also discuss the radion dynamics and braneworld black holes 
for the close limit in our formulation.   
\end{abstract}
\vskip2pc]


\vskip1cm


\section{Introduction}

The purpose of this paper is to write down the effective theory for two branes system 
at the closed limit.  This kind of system is a situation just before the collision of 
two branes. Since the brane is a key object in superstring theory and braneworld 
scenarios\cite{early,RSI,RSII,old}, the collision 
process of branes definitely could be an important and fundamental 
process.  In a braneworld the brane-brane collision could 
provide us a new picture of the creation 
of the big-bang universe\cite{collision,debate,collision2,KSS}. A brane-bubble 
collision also will be interesting because it might give us a 
new scenario of the inflation 
and reheating process\cite{BB}. Thus it is worth deriving an effective theory on the brane for 
close limit of two branes.

To derive an effective theory on the brane, we must solve the bulk spacetime. In the 
ordinary Kaluza-Klein theory it was a trivial task. In the braneworld, it is quite 
non-trivial because the gravity on the brane is non-trivially coupled to the gravitational
fields in the bulk. Recently it has been reported that we can do so for a two branes system 
if we employ the derivative expansion scheme\cite{Tess2,Kanno,Kanno2}, that is, 
long wave approximations. 
The two junction conditions on the branes gives us sufficient boundary conditions to determine the
bulk geometry. Thus, the bulk geometry could be uniquely determined. Finally, a low energy effective 
theory can be derived (See also Refs. \cite{Toby} and \cite{Csaki} for earlier works and 
Refs. \cite{Gen,Kazuya,Other,Kudoh2} for the other issues.). In this paper, on the other hand, 
we point out that we can apply the Taylor expansion to the closed limit of two branes. 
In this method, the high energy effects 
can be easily seen, which are difficult to deal with in low energy approximations. The price of it
is that the separation between two branes should be small. 

In this paper we will use a toy model to concentrate on the dealing of the 
gravitational effect and dynamics for two branes system of the Randall-Sundrum 
model\cite{RSI} where the bulk stress energy is a cosmological constant. 
The brane will be treated as the thin domain wall. 
In this approximation, we will be able to use the Israel junction 
formalism\cite{Israel}.  
However a problem comes out here. 
From the derivation it is obvious that the Israel junction condition cannot be applied 
to just a moment of the collision 
of two branes with $Z_2$-{\it symmetry}. That is, under $Z_2$-symmetry, the size of the extra 
dimension shrinks to zero at the collision.  
So it is fair to say that our present analysis cannot 
treat the collision in Randall-Sundrum type models. Our formalism can be applied
just before the collision of two branes. In order to treat the collision, 
we need the interaction between branes, that is, the brane field action which will be 
reduced to something such as the Born-Infeld action\cite{BI} at the thin wall limit. 
This is, however, beyond the scope of present study.

In this paper we will derive the effective theory describing the close limit of two branes 
and discuss the radion dynamics, which expresses the evolution of the brane separation, 
and its relation to the bulk geometry. We also give an insight into the 
braneworld black holes. Our present 
study can be regarded as the generalization of that by Khoury and Zhang\cite{Justin} who 
considered the Friedman brane cosmology in an empty bulk and discussed the meaning of  
dark radiation\cite{DR}. We will work in the covariant curvature 
formalism\cite{Tess,Roy} which is most useful to keep the four dimensional 
covariance and obtain the full view. 

The rest of this paper is organised as follows. In Sec. II, 
we review the covariant curvature formalism. In Sec. III, 
we solve the bulk geometry by Taylor expansion and then derive the 
four dimensional effective theory at the close limit.  In Sec.  
III the radion dynamics and the braneworld black holes will be 
addressed. Finally we will give the summary and discussion in Sec. IV. 

\section{Covariant curvature formalism}

In this section we review the covariant curvature formalism developed in Refs. \cite{Tess,Roy}.
Note that the central issue is how to solve the bulk seriously. 
Without this step we cannot obtain a correct effective theory on the brane. 

We employ the metric form
%
\begin{eqnarray}
ds^2=e^{2\phi(x)}dy^2+q_{\mu\nu}(y,x)dx^\mu dx^\nu.\label{metric}
\end{eqnarray}
%
In the above it is supposed that 
the positive and negative tension branes are located at $y=0$ and $y=y_0$, 
respectively. The proper distance between two branes is given by $d_0(x)=\int_0^{y_0}
e^{\phi (x)} dy=y_0 e^{\phi (x)}$. 
$q_{\mu\nu}(y,x)$ is the induced metric of $y=$const hypersurfaces. At the 
close limit of two branes we expect that the coordinate used 
in the metric of Eq. (\ref{metric}) does not break down for a wide class of the 
braneworld spacetimes. 
 
From the Gauss-Codacci equations, we have two key equations
%
\begin{eqnarray}
{}^{(4)}G^\mu_\nu & = & \frac{3}{\ell^2} \delta^\mu_\nu+KK^\mu_\nu-K^\mu_\alpha K^\alpha_{\nu}
\nonumber \\
& & -\frac{1}{2}\delta^\mu_\nu (K^2-K^\alpha_\beta K^\beta_\alpha)
-E^\mu_\nu \label{Gauss}
\end{eqnarray}
%
and
%
\begin{eqnarray}
D_\mu K^\mu_\nu-D_\nu K =0,
\end{eqnarray}
%
where $D_\mu$ is the covariant derivative with respect to $q_{\mu\nu}$. $\ell$ is the bulk  
curvature radius. 
${}^{(4)}G_{\mu\nu}$ is the four-dimensional Einstein tensor with respect to 
$q_{\mu\nu}$ and 
$K_{\mu\nu}$ is the extrinsic curvature of $y=$constant hypersurfaces defined by 
%
\begin{eqnarray}
K_{\mu\nu}=\frac{1}{2}\mbox \pounds_n q_{\mu\nu}=\nabla_\mu n_\nu+n_\mu D_\nu \phi,
\end{eqnarray}
%
where $n=e^{-\phi}\partial_y$. Here note that $a^\mu= n^\nu \nabla_\nu n^\mu=-D^\mu \phi (x)$.  
$E_{\mu\nu}$ is a part of the projected Weyl tensor defined by 
%
\begin{eqnarray}
E^\mu_\nu & = & {}^{(5)}C_{\mu\alpha\nu\beta}n^\alpha n^\beta \nonumber \\ 
& = &  -D^\mu D_\nu \phi - D^\mu \phi D_\nu \phi - \mbox \pounds_n K^\mu_\nu-K^\mu_\alpha K^\alpha_\nu
 \nonumber \\
& & +\frac{1}{\ell^2}\delta^\mu_\nu, \label{defE}
\end{eqnarray}
%
where ${}^{(5)}C_{\mu\nu\alpha\beta}$ is the five-dimensional Weyl tensor. 

The junction conditions on the branes give us 
the relation between the extrinsic curvatures and 
the energy-momentum tensor on the branes:
%
\begin{eqnarray}
[K^\mu_\nu-\delta^\mu_\nu K ]_{y=0}=-\frac{\kappa^2}{2}
\biggl(-\sigma_1 \delta^\mu_\nu+T^{\mu}_{1~\nu} \biggr) 
\end{eqnarray}
%
and
%
\begin{eqnarray}
[K^\mu_\nu-\delta^\mu_\nu K ]_{y=y_0}=\frac{\kappa^2}{2}
\biggl(-\sigma_2 \delta^\mu_\nu+T^{\mu}_{2~\nu} \biggr). 
\end{eqnarray}
%
$T^\mu_{1~\nu}$ and $T^\mu_{2~\nu}$ are the energy-momentum tensor localised on the 
positive and negative branes. $\sigma_1$ and $\sigma_2$ are the brane tensions. 
If one substitutes the above conditions to Eq. (\ref{Gauss}), 
we might be able to derive the Einstein equation on the brane\cite{Tess}. 
However, it turns out that $E_{\mu\nu}$ is not still unknown, because it is a five dimensional quantity. 
For the single brane $E_{\mu\nu}$ is identical to the contribution from 
the Kaluza-Klein modes in the linealised 
theory\cite{Tama} and it will be higher order correction at large distance, that is, $E_{\mu\nu}$ is 
negligible at low energy in the linearlised theory. 
For the two branes system, on the other hand, $E_{\mu\nu}$ is not negligible as seen later. 
Anycase, we need the evolutional equation for $E_{\mu\nu}$ to the bulk. 

To evaluate $E_{\mu\nu}$ in the bulk, we derive its evolutional equation. The result is 
given by 
%
\begin{eqnarray}
& & \mbox \pounds_n E_{\alpha\beta} =D^\mu B_{\mu (\alpha\beta)}+K^{\mu\nu}{}^{(4)}C_{\mu\alpha\nu\beta}
+4K^\mu_{(\alpha}E_{\beta) \mu} \nonumber \\ 
& &~~~~~~ -\frac{3}{2}KE_{\alpha\beta}-\frac{1}{2}q_{\alpha\beta}K^{\mu\nu}E_{\mu\nu}
+2D^\beta \phi B_{\beta (\mu\nu)}
 \nonumber \\
& & 
~~~~~~+2\tilde K^\mu_\alpha \tilde K_{\mu\nu} \tilde K^\nu_\beta-\frac{7}{6}\tilde K_{\mu\nu}
\tilde K^{\mu\nu}\tilde K_{\alpha\beta} \nonumber \\ 
& &~~~~~~ -\frac{1}{2}q_{\mu\nu}\tilde K_{\alpha\beta}
\tilde K^\alpha_\rho \tilde K^{\rho\beta}, \label{evoE}
\end{eqnarray}
%
where $B_{\mu\nu\alpha}=q_\mu^\rho q_\nu^\sigma {}^{(5)}C_{\rho\sigma\alpha\beta}n^\beta$ and 
$\tilde K_{\mu\nu}=K_{\mu\nu}-\frac{1}{4}q_{\mu\nu}K$. 
Since the right-hand side contains $B_{\mu\nu\alpha}$, $K_{\mu\nu}$, and 
${}^{(4)}C_{\mu\nu\alpha\beta}$, we need their evolutional equations too. 
After a long calculation we obtain  
%
\begin{eqnarray}
& & \mbox \pounds_n {}^{(4)}R_{\mu\nu\alpha\beta}+2{}^{(4)}R_{\mu\nu\rho [\alpha}K^\rho_{\beta]}
+2D_{[\mu}B_{|\alpha\beta |\nu]} \nonumber \\
& &~~~ +2(D_\mu D_{[\alpha} \phi+ D_\mu \phi D_{[\alpha} \phi)K_{\beta]\nu} \nonumber \\
& & ~~~-2(D_{\nu}D_{[\alpha} \phi -D_\nu \phi D_{[\alpha}\phi )K_{\beta ] \mu} 
\nonumber \\
& &~~~ -2B_{\alpha\beta [\mu} D_{\nu ]}\phi -2 B_{\mu\nu [\alpha} D_{\beta ]} \phi=0,
\end{eqnarray}
%
%
\begin{eqnarray}
& & \mbox \pounds_n  B_{\mu\nu\alpha}+2D_{[\mu} E_{\nu ]\alpha}
+2D_{[\mu}\phi E_{\nu]\alpha} \nonumber \\
& & ~~~- B_{\mu\nu\beta}K^\beta_\alpha+2 B_{\alpha\beta [\mu}
K_{\nu]}^\beta \nonumber \\
& &~~~ +({}^{(4)}R_{\mu\nu\alpha\beta}-K_{\mu\alpha}K_{\nu\beta}
+K_{\mu\beta}K_{\nu\alpha})D^\beta \phi =0,
\end{eqnarray}
%
and
%
\begin{eqnarray}
e^{-\phi}\partial_y K^\mu_\nu & = & -D^\mu D_\nu \phi-D^\mu \phi D_\nu \phi -K^\mu_\alpha 
K^\alpha_\nu+\frac{1}{\ell^2}\delta^\mu_\nu \nonumber \\
& & -E^\mu_\nu. \label{evoK}
\end{eqnarray}
%
Equation (\ref{evoK}) is a rearrangement of Eq. (\ref{defE}). 

The junction condition directly implies the boundary condition on the branes for 
$K_{\mu\nu}$ and $B_{\mu\nu\alpha}$ because of 
%
\begin{eqnarray}
B_{\mu\nu\alpha}=2D_{[\mu} K_{\nu]\alpha}.\label{weylb}
\end{eqnarray}
%

For the discussion in the next section, we write down the second derivative of 
$K^\mu_\nu $:
%
\begin{eqnarray}
\partial_y^2 K^\mu_\nu & = &  e^{2\phi}
\biggl[ K^\alpha_\nu D^\mu \phi D_\alpha \phi+ K^\mu_\alpha D_\nu 
\phi D^\alpha \phi
\nonumber \\
& & +3 K^\mu_\alpha (D^\alpha D_\nu \phi +D^\alpha \phi 
D_\nu \phi) \nonumber \\
& & +K^\alpha_\nu (D^\mu D_\alpha \phi +D^\mu \phi D_\alpha \phi)
-K^\alpha_\nu E^\mu_\alpha+K^\mu_\alpha E^\alpha_\nu \nonumber \\
& & +\frac{3}{2}KE^\mu_\nu+\frac{1}{2}\delta^\mu_\nu K^\alpha_\beta E^\beta_\alpha+
D_{\alpha} \phi D^{\mu} K^{\alpha}_{\nu}-K^{\mu}_{\nu} (D \phi)^2 \nonumber\\
& & + B_{\nu\alpha}^{~~\mu}
D^\alpha\phi-D^\alpha \phi (-B^\mu_{~\alpha\nu}+B_{\alpha\nu}^{~~\mu})
 +\chi^\mu_\nu \biggr] \label{evoK2}
\end{eqnarray}
%
where $\chi^\mu_\nu$ is defined by 
%
\begin{eqnarray}
\chi^\mu_\nu & =&-\frac{1}{2}(-D^\alpha B^\mu_{~\alpha\nu}+D^\alpha B_{\alpha\nu}^{~~\mu})
-K^{\alpha\beta} {}^{(4)} C^\mu_{~\alpha\nu\beta}
\nonumber \\
& & +2K^\mu_\alpha K^\alpha_\beta K^\beta_\nu 
-\frac{2}{\ell^2}K^\mu_\nu -2 \tilde K^\mu_\alpha \tilde K^\alpha_\beta \tilde K^\beta_\nu
\nonumber \\
& & +\frac{7}{6}\tilde K^\mu_\nu \tilde K^\alpha_\beta \tilde K^\beta_\alpha 
+\frac{1}{2}\delta^\mu_\nu \tilde K^\alpha_\beta \tilde K^\rho_\alpha \tilde K^\beta_\rho.
\end{eqnarray}
%

See Ref. \cite{Tess2} for the derivation of the low energy effective 
theory by using the present covariant formalism. In Ref. \cite{Tess2} we employed  
the derivative expansion scheme which corresponds to the long-wave 
approximation, that is, the typical scale on the brane is much larger than the 
bulk curvature scale. In other words, 
the energy scale of the energy momentum tensor localized on the branes should 
be sufficiently lower than the tension of the branes. As said before, in this 
paper, we will focus on the close limit and employ the Taylor expansion 
assuming that the distance between two branes 
$d_0$ is much shorter than the bulk curvature scale $\ell$. In this method, 
it is not necessary to take the Randall-Sundrum tuning of the tensions and low energy 
approximation $\kappa^2T^{\mu}_{\nu}/\sigma \ll 1$.

\section{Effective theory at closed limit}

In this section, we solve the bulk geometry using the Taylor expansion in 
the covariant curvature formalism. What we will do is the evaluation of $E_{\mu\nu}$, that is, 
the writing down of $E_{\mu\nu}$ in terms of the four dimensional quantities on the branes. 
Then we derive the effective equation on 
the branes. This theory can be applicable to the non-linear regime as long as we 
consider the close limit. 

The point of the derivation is simple. We consider the Taylor expansion of the 
extrinsic curvature, $K^\mu_\nu (y_0,x)=K^\mu_\nu (0,x)+\partial_y K^\mu_\nu (0,x)y_0+\cdots$. 
The derivative of the extrinsic curvature $\partial_y K_{\mu\nu}$ 
contains the bulk informations($E_{\mu\nu}$), and then the junction condition 
determines them in terms 
of the four dimensional quantity. This is the specialty in two brane systems.  
If one considers a single-brane system, we need a boundary condition in the 
bulk somewhere such as Cauchy horizon. 

In  this paper we  will think of the positive tension brane. The extension of the present discussion to the 
negative tension brane is very easy. 
We begin with the Taylor expansion of the extrinsic curvature $K^\mu_\nu$ 
around the positive tension brane
%
\begin{eqnarray}
K^\mu_\nu (y_0,x) & = & K^\mu_\nu (0,x)+\partial_y K^\mu_\nu (0,x)y_0 \nonumber \\
& & +\frac{1}{2}\partial_y^2 K^\mu_\nu (0,x) y_0^2 +O(y^3). \label{Tay1}
\end{eqnarray}
%
Using Eqs (\ref{evoK}) and (\ref{evoK2}), Eq. (\ref{Tay1}) becomes  
%
\begin{eqnarray}
K^\mu_\nu(y_0,x) & = &   K^\mu_\nu (0,x)-d_0 
\biggl[ \frac{D^\mu D_\nu d_0}{d_0}  +K^\mu_\alpha 
K^\alpha_\nu   +E^\mu_\nu \nonumber \\
& & -\frac{1}{\ell^2}\delta^\mu_\nu \biggr]
+\frac{1}{2}d_0^2
\biggl[ \chi^\mu_\nu +3\frac{D^\alpha D_\nu d_0}{d_0} K^\mu_\alpha 
\nonumber \\
& & + \frac{D_\alpha D^\mu d_0}{d_0}K^\alpha_{\nu} +
K^{\alpha}_{\nu} \frac{D^{\mu} d_0 D_{\alpha} d_0}{d_0^2}
\nonumber \\
& &
+K_{\alpha}^{\mu} \frac{D_{\nu} d_0 D^{\alpha} d_0}{d_0^2}
-K^{\mu}_{\nu} \left(\frac{D d_0}{d_0} \right)^2 \nonumber \\
& & +\frac{D^{\alpha} d_0}{d_0}\biggl[
 D^{\mu} K_{\alpha\nu} +B^{\mu}_{~\alpha \nu}+2B_{\nu\alpha}^{~~~\mu}\biggr] \nonumber\\
& & - K^\alpha_\nu E^\mu_\alpha +K^\mu_\alpha E^\alpha_\nu 
+\frac{3}{2}KE^\mu_\nu+\frac{1}{2}\delta^\mu_\nu E^\alpha_\beta K^\beta_\alpha
\biggr]\nonumber \\
& & +O(d_0^3), \label{Taylor1}
\end{eqnarray}
%
where $K^\mu_\nu $ and $E^\mu_\nu $ in the right-hand side is evaluated at $y=0$. 

Let us regard Eq. (\ref{Taylor1}) as the equation for $E^\mu_\nu$.
To solve the equation, we assume that two branes are close enough 
to satisfy the conditions 
\begin{equation}
d_0/\ell,\quad \kappa^2 T_1 d_0, \quad \kappa^2 T_2 d_0 <1.
\end{equation}
Then we expand $E^\mu_\nu$ as 
%
\begin{eqnarray}
E^\mu_\nu =\stac{(-1)}{E^\mu_\nu} \frac{1}{d_0}+\stac{(0)}{E^\mu_\nu} +O(d_0).
\end{eqnarray}
%
Substituting the above into Eq. (\ref{Taylor1}) and solving iteratively we see that   
$\stac{(-1)}{E^\mu_\nu}$ and $\stac{(0)}{E^\mu_\nu}$ has the following form: 
%
\begin{eqnarray}
-\stac{(-1)}{E^\mu_\nu} = K^\mu_\nu (y_0,x)-K^\mu_\nu +D^\mu D_\nu d_0 \label{Just}
\end{eqnarray}
%
and
%
\begin{eqnarray}
-\stac{(0)}{E^\mu_\nu} & =&
K^\mu_\alpha K^\alpha_\nu -\frac{1}{\ell^2}\delta^\mu_\nu 
 -\frac{1}{2}\biggl( -K^\alpha_\nu \stac{(-1)}{E^\mu_\alpha}+K^\mu_\alpha \stac{(-1)}{E^\alpha_\nu}
\nonumber \\
& & +\frac{3}{2}K \stac{(-1)}{E^\mu_\nu}
+\frac{1}{2}\delta^\mu_\nu K^\alpha_\beta \stac{(-1)}{E^\beta_\alpha}  \biggr) \nonumber \\  
& & -\frac{1}{2}(3K^\mu_\alpha D^\alpha D_\nu d_0  +K^\alpha_\nu D_\alpha D^\mu d_0) \nonumber \\
& & -\frac{D^\mu d_0 D_\alpha d_0}{2d_0} K^\alpha_\nu 
 -\frac{D_\mu d_0 D^\alpha d_0}{2d_0} K_\alpha^\mu 
\nonumber\\
& &+\frac{1}{2}K^{\mu}_{\nu} 
\frac{(D d_0)^2}{d_0} -\frac{1}{2} D_{\alpha} d_0D^{\mu} K^{\alpha}_{\nu} \nonumber\\
& & +\frac{1}{2}D^\alpha d_0 (2B_{\alpha\nu}^{~~~\mu}-B^\mu_{~\alpha\nu}) .
\end{eqnarray}
%
In the above we supposed that $D^2 d_0$ is the same order as $K$. 
We should notice that it is not
necessary to impose low energy conditions $T /\sigma \sim 
\kappa^2 T \ell <1$. What we imposed here is $\kappa^2 T d_0 <1$. 

From the fact that $E^\mu_\nu$ is traceless we can have the equation 
for $d_0$
%
\begin{eqnarray}
& & \frac{1}{d_0}\biggl( 1+\frac{d_0}{\ell_1}\biggr)D^2 d_0-\frac{4}{d_0}\biggl(\frac{1}{\ell_2}
-\frac{1}{\ell_1} \biggr)+4 \biggl(\frac{1}{\ell_1 \ell_2}-\frac{1}{\ell^2} \biggr) \nonumber \\
& & ~~-\frac{\kappa^2}{6d_0}
\biggl[ \biggl( 1+ \frac{d_0}{\ell_2}\biggr)T_1+\biggl(1-\frac{d_0}{\ell_1} \biggr)T_2 \biggr] \nonumber \\
& & ~~+\frac{\kappa^2}{2}
\biggl(T^\alpha_{1~\beta}-\frac{1}{3}\delta^\alpha_\beta T_1 \biggr)D^\beta D_\alpha d_0
\nonumber \\
& & ~~+\frac{\kappa^2}{2}\biggl(T^\alpha_{1~\beta}-\frac{1}{6}\delta^\alpha_\beta T_1 \biggr)
\frac{D^\beta d_0 D_\alpha d_0}{d_0} \nonumber\\
& & ~~- \frac{(Dd_0)^2}{\ell_1 d_0}-\frac{\kappa^4}{4}
\biggl(  T^\alpha_{1~\beta}T^\beta_{2~\alpha}-\frac{2}{9}T_1 T_2 \biggr) \nonumber \\
& & ~~-\frac{\kappa^2}{12}D^\alpha d_0 D_\alpha T_1 =0,
\label{limit2}
\end{eqnarray}
%
where we defined 
%
\begin{eqnarray}
\frac{1}{\ell_1}= \frac{1}{6}\kappa^2 \sigma_1~~{\rm and}~~\frac{1}{\ell_2}=-\frac{1}{6}\kappa^2 \sigma_2.
\end{eqnarray}
%

Since we write down $E^\mu_\nu$ in terms of four dimensional quantities, 
we can derive the effective gravitational equation on the brane
%
\begin{eqnarray}
{}^{(4)}G^\mu_\nu & = &  6 \biggl(\frac{1}{\ell^2}-\frac{1}{\ell_1^2} \biggr) \delta^\mu_\nu
-\frac{9}{\ell_1}\biggl( \frac{1}{\ell_2}-\frac{1}{\ell_1}\biggr)\delta^\mu_\nu \nonumber \\
& & +\frac{3}{d_0}\left (\frac{1}{\ell_2}-\frac{1}{\ell_1} \right) \delta^{\mu}_{\nu}
    +\frac{\kappa^2}{2d_0}(T^\mu_{1~\nu}+T^\mu_{2~\nu}) \nonumber\\
& & +\frac{3\kappa^2}{2}\biggl(\frac{1}{\ell_2}-\frac{1}{\ell_1} \biggr) T_1 \delta^\mu_\nu
    +\frac{\kappa^2}{2\ell_1}(T^\mu_{1~\nu}-3T^\mu_{2~\nu}) \nonumber \\
& & +\biggl( 1-\frac{d_0}{\ell_1}\biggr) \frac{D^\mu D_\nu d_0 -\delta^\mu_\nu D^2 d_0}{d_0} \nonumber \\
& & +\frac{2D^\mu d_0 D_\nu d_0+\delta^\mu_\nu (Dd_0)^2}{2\ell_1 d_0}\nonumber \\
& & +\kappa^2 \biggl[ -\frac{5}{24}T_1 D^\mu D_\nu d_0 -\frac{1}{6}\delta^\mu_\nu T_1 D^2 d_0
\nonumber \\
& & +\frac{1}{2}(T^\mu_{1~\alpha}D^\alpha D_\nu d_0+T^\alpha_{1~\nu}D_\alpha D^\mu d_0) 
\nonumber \\
& & -\frac{5}{8}\delta^\mu_\nu T^\alpha_{1~\beta}D^\beta D_\alpha d_0 \nonumber \\
& &+\frac{1}{4}T^\alpha_{1~\nu} \frac{D^\mu d_0 D_\alpha d_0}{d_0}
+\frac{1}{4}T_{1~\alpha}^{\mu} \frac{D_\nu d_0 D^\alpha d_0}{d_0}
\nonumber \\
& &-\frac{1}{6}T_1 \frac{D^\mu d_0 D_\nu d_0}{d_0}+\frac{1}{6}\delta^\mu_\nu T_1 \frac{(Dd_0)^2}{d_0}
\nonumber \\
& & -\frac{1}{4}T^\mu_{1~\nu} \frac{(D d_0)^2}{d_0}-\frac{1}{2}\delta^\mu_\nu T^\alpha_{1~\beta} 
\frac{D^\beta d_0 D_\alpha d_0}{d_0} \nonumber \\
& & -\frac{1}{4} D^\alpha d_0 ( 3 D_{\alpha} T_{1~\nu}^{\mu}
-2 D_{\nu} T_{1~\alpha}^{\mu}-2 D^{\mu} T_{1~\nu \alpha}) \nonumber\\
& & +\frac{1}{12}(4 \delta^{\mu}_{\nu} D^{\alpha} d_0 D_{\alpha} T_1
 -2 D^{\mu} d_0 D_{\nu} T_1 \nonumber \\
& & -2 D_{\nu} d_0 D^{\mu} T_1)  
\biggr] \nonumber\\
& & +\kappa^4 \biggl[ 
-\frac{1}{12}T_1 T^\mu_{1~\nu}+\frac{1}{24}\delta^\mu_\nu T_1^2-\frac{1}{8}
\delta^\mu_\nu T^\alpha_{1~\beta}T^\beta_{1~\alpha} \nonumber \\
& & +\frac{1}{8}
(T^\alpha_{1~\nu}T^\mu_{2~\alpha}-T^\alpha_{2~\nu}T^\mu_{1~\alpha})
+\frac{1}{16}T_1 (T^\mu_{1~\nu}+T^\mu_{2~\nu})
\nonumber \\
& & +\frac{3}{16}\delta^\mu_\nu T^\alpha_{1~\beta}(T^\beta_{1~\alpha}+ T^\beta_{2~\alpha} )
\biggr].
\end{eqnarray}
%
Here note that we derived the above equation under the condition of 
$d_0/\ell,\quad \kappa^2 T_1 d_0, \quad \kappa^2 T_2 d_0 <1$.  

The coupling of the matter with the radion is complicated. 
In the low energy limit such terms 
can be omitted and after taking of Randall-Sundrum tuning of tensions 
$\ell=\ell_1=\ell_2$, the gravitational equation is reduced to  
%
\begin{eqnarray}
{}^{(4)}G^\mu_\nu & = &  
\frac{\kappa^2}{2d_0}(T^\mu_{1~\nu}+T^\mu_{2~\nu})+\frac{\kappa^2}{2\ell}(T^\mu_{1~\nu}-3T^\mu_{2~\nu}) 
\nonumber \\
& & +\biggl( 1-\frac{d_0}{\ell}\biggr) \frac{D^\mu D_\nu d_0 -\delta^\mu_\nu D^2 d_0}{d_0}
\nonumber \\
& & 
+\frac{2 D^\mu d_0 D_\nu d_0+\delta^\mu_\nu (Dd_0)^2}{2\ell d_0}.
\end{eqnarray}
%
Here reminded that 
the first term is much bigger than the second term in the first line of 
the right-hand side. It is easy to check that the above is identical to 
the close limit of the low energy effective theory obtained in Refs. 
\cite{Tess2,Kanno}.

\section{Applications}

In this section we consider the applications of our formalism: 
the radion dynamics and the braneworld black holes. 

\subsection{Radion dynamics and bulk geometry} 

First, let us remember the work of Khoury and Zhang \cite{Justin}. They investigated the modified 
Friedman equation on the brane in empty bulk and observed that $E_{00}$ is proportional to 
the energy density on the brane. It corresponds to Eq (\ref{Just}) in our 
general analysis of the close limit because we can neglect the 
effect of bulk curvature for $d_0 \ll \ell$. At the leading order of 
$d_0/\ell, d_0 \kappa^2 T$, the effective equation on the positive tension brane becomes
%
\begin{eqnarray}
G^{\mu}_{\nu} & = & -E^{\mu}_{\nu}, \nonumber\\
-E^\mu_\nu & = &  \frac{\kappa^2}{2d_0}(T^\mu_{1~\nu}+T^\mu_{2~\nu})
+\frac{D^\mu D_\nu d_0 -\delta^\mu_\nu D^2 d_0}{d_0}, \nonumber\\ 
D^2 d_0 &=& \frac{\kappa^2}{6} (T_1+T_2).
\end{eqnarray}
%
From these effective equations, it is easy to derive the modified Friedmann equation
on the brane
%
\begin{equation}
H^2=-H \frac{\dot{d_0}}{d_0}+ \frac{\kappa^2(\rho_1+\rho_2)}{6 d_0},
\label{Friedmann}
\end{equation}
%
where $H$ is the Hubble parameter 
on the positive tension brane. $\rho_1$ and $\rho_2$ are energy densities 
on positive tension brane and negative tension brane, respectively. 
This agrees with Khoury and Zhang's equation. 
From the evolution equation for radion and the conservation law for the 
energy-momentum tensor, the right-hand side of Eq. (\ref{Friedmann}) can 
be rewritten as 
%
\begin{equation}
H^2=C a^{-4},
\end{equation}
%
where $a$ is the scale factor on the positive tension brane and 
$C$ is the integration constant of the radion $d_0$. Thus we can 
understand the behavior of dark radiation in terms of the dynamics of the radion and 
that the bulk geometry is related to the initial condition for the radion. 
Hence the present formalism gives us 
a straightforward way to see what they observed. 

\subsection{Radion dynamics for deSitter-like branes}

Next we discuss the radion dynamics for the deSitter-like branes. For simplicity we consider the 
vacuum energy cases
%
\begin{eqnarray}
\ell=\ell_1=\ell_2
\end{eqnarray}
%
and
%
\begin{eqnarray}
T^\mu_{1~\nu}=-\lambda_1 \delta^\mu_\nu~~T^\mu_{2~\nu}=-\lambda_2 \delta^\mu_\nu. 
\end{eqnarray}
%
Then the branes will be deSitter spacetimes if the maximum symmetry is imposed. 
But, we do not restrict ourself to the consideration on the exact deSitter branes. 
In this case the equation for the radion is extremely simplified as 
%
\begin{eqnarray}
& & \biggl(1+\frac{d_0}{L} \biggr) \frac{D^2 d_0}{d_0}
-\frac{(D d_0)^2}{Ld_0} -\frac{\kappa^4}{9}\lambda_1 \lambda_2\nonumber \\
& & ~~+\frac{2\kappa^2}{3d_0}\biggl[ \biggl(1+\frac{d_0}{\ell} \biggr)\lambda_1 
+\biggl(1-\frac{d_0}{\ell} \biggr) \lambda_2 \biggr]=0,
\end{eqnarray}
%
where $1/L=1/\ell + \kappa^2 \lambda_1 /6$. Defining $ \varphi 
= d_0/\ell $($\ll 1$) and the conformaly transformed metric $g_{\mu\nu}=(1+\varphi)^{-1}q_{\mu\nu}$, 
the equation becomes 
%
\begin{eqnarray}
D^2_g \varphi -V'(\varphi) = 0,
\end{eqnarray}
%
where 
%
\begin{eqnarray}
V(\varphi) \simeq   \alpha \varphi +\beta \varphi^2 +\cdots.
\end{eqnarray}
%
The constants $\alpha$ and $\beta$ are defined by 
%
\begin{eqnarray}
\alpha = -\frac{2\kappa^2}{3L}(\lambda_1+\lambda_2)
\end{eqnarray}
%
and 
%
\begin{eqnarray}
\beta = -\frac{\kappa^2}{3\ell}(\lambda_1 -\lambda_2)+\frac{\kappa^4}{18}\lambda_1 
\lambda_2.
\end{eqnarray}
%

The close limit corresponds to $\varphi  \to 0$ limit. At the very close 
limit, the potential is approximated by 
%
\begin{eqnarray}
V(\varphi) \simeq  \alpha \varphi.
\end{eqnarray}
%
So the brane could collide if $\alpha <0$ due to the attractive force. 
If $\alpha >0$ the repulsive force acts between two branes. This 
feature is same as the result in Ref. \cite{KSS}. The term 
proportional to $\beta$ describes the correction in the next order. 
We should emphasize that 
high-energy terms such as $\kappa^2 \lambda_1 \lambda_2$ can be included
in our method. In the case of $\alpha =0$, 
the next order should be taken into account and the potential becomes 
%
\begin{eqnarray}
V(\varphi) \simeq -\frac{2\kappa^2}{3\ell}\lambda_1 \biggl( 
1+\frac{1}{12}\kappa^2 \lambda_1 \ell \biggr) \varphi^2.
\end{eqnarray}
%
From the above we can see that the repulsive force will be induced between two branes when 
$\lambda_1 >0$. 

In general the potential at the close limit is given by 
%
\begin{eqnarray}
V(\varphi) \simeq  \frac{\kappa^2}{6L}(T_1+T_2)\varphi+\cdots.
\end{eqnarray}
%
Thus the attractive force  acts if $T_1+T_2>0$ and the repulsive force if 
$T_1+T_2 <0$. If one thinks of the dust universe, the repulsive force does. 

\subsection{An implication into black holes}

Let us finally consider brane world black holes in the case of $\ell=\ell_1=\ell_2$ and $T^\mu_{1~\nu}
=T^\mu_{2~\nu}=0$. In this case the equation for the radion and the gravitational equation 
become 
%
\begin{eqnarray}
\frac{1}{d_0}\biggl(1+\frac{d_0}{\ell} \biggr)D^2 d_0 -\frac{(Dd_0)^2}{\ell d_0}=0
\end{eqnarray}
%
and
%
\begin{eqnarray}
{}^{(4)}G^\mu_\nu &  = &  \biggl(1-\frac{d_0}{\ell} \biggr)\frac{D^\mu D_\nu d_0 -\delta^\mu_\nu 
D^2 d_0}{d_0} \nonumber \\
& & +\frac{2D^\mu d_0 D_\nu d_0+\delta^\mu_\nu (D d_0)^2}{2\ell d_0}.
\end{eqnarray}
%
Up to this order the higher order correction disappears because they are written 
by the energy-momentum tensor on the branes. The above system is equivalent to  
the scalar-tensor theory. In the present case it is easy to see that the 
static and spherical solution on the brane is only Schwarzshild spacetimes 
due to the Bekenstein theorem\cite{nohair}. At the close limit, thus, the black holes 
in the bulk are described by the adS black string solution\cite{BH}. However, if 
the brane distance increase, it is natural to expect that there should be black holes 
solution confined on the each branes. That is, there might be a phase 
transition from black holes to the adS black string\cite{BHBS}. Thus we hope we will be able to 
see this transition using our formalism. This expectation might be supported because 
the higher order corrections than that we considered here will contain the terms like 
$D^\alpha {}^{(4)}R_{\mu\nu\alpha\beta} D^\beta \phi$ and ${}^{(4)}R_{\mu\alpha\nu\beta}
{}^{(4)}R^{\alpha\beta}$ and then the 
modified solution from the Schwarzshild could be emerged such as the situation 
in the superstring theory\cite{STBH}.

\section{Summary and discussion}

In this paper we deriveed the effective theory at the close limit of two branes 
and discussed the radion dynamics/braneworld black holes. Our effective 
theory is complimentary with the 
low energy effective theory\cite{Tess2,Kanno} and can be regarded as 
the generalisation of the Khoury and Zhang's recent work\cite{Justin}.

Finally we should stress on the limitation of our formulation. (i)To determine the 
bulk spacetime we used the Israel's junction condition at the beginning. 
Obviously the derivation 
of the junction condition will be not correct at the just moment of 
collision of two branes. 
Our formalism is correct before the collision. (ii)In addition, we tacitly supposed 
that the coordinates to the extra dimensions works very well. (iii)In order to treat 
the collision we must think of the thick wall where 
the brane is described by a field theory and the brane-brane interactions is also 
given. Since the effective theory describing the thick wall might be reformulated 
in terms of the noncommutative geometry\cite{NC}, the brane collision might be also 
treated in the effective theory with the noncommutative geometric aspects. 

\section*{Acknowledgments}

We would like to thank Shoko Hayakawa, Daisuke Ida, Takashi Torii, and Toby 
Wiseman for 
fruitful discussions. This work is inspired by unpublished study\cite{unpub} 
on single brane models. The work of TS was supported by Grant-in-Aid for Scientific
Research from Ministry of Education, Science, Sports and Culture of
Japan(No. 13135208, No.14740155 and No.14102004). The work of KK were 
supported by the JSPS. 


\end{document}